# COMPARISON BETWEEN SPIN AND ROTATION PROPERTIES OF LORENTZ EINSTEIN AND REFLECTION SYMMETRIC TRANSFORMATIONS

Mushfiq Ahmad
Department of Physics, Rajshahi University, Rajshahi, Bangladesh.
E-mail: mushfiqahmad@ru.ac.bd

M. Shah Alam
Department of Physics, Shah Jalal University of Science and Technology, Sylhet, Bangladesh.
E-mail: salam@sust.edu

M.O.G. Talukder
Department of Applied Physics and Electronics, Rajshahi University, Rajshahi, Bangladesh.
E-mail: ogtalukder@ru.ac.bd

## Abstract

We have shown that reflection symmetric transformation is Lorentz invariant; it is also associative. We have also shown that reflection symmetric sum of vectors has a spin-like term comparable to the spin of Dirac electron. As a consequence of reflection symmetry we have found that the sum is bounded. This corresponds to Einstein's postulate.

Key words: Reflection symmetry, Lorentz invariance, Spin, Associativity.

## 1. Introduction

We have defined a reflection symmetry sum $\hat{+}$ ( + with a cap ^) of vectors **A** and **B** as [1].

$$\mathbf{A} \hat{+} \mathbf{B} = \frac{\mathbf{A} + \mathbf{B} + i\mathbf{AxB}}{\mathbf{1} + \mathbf{A.B}} \tag{1}$$

$\mathbf{A'}$ will be called a reciprocal of $\mathbf{A}$ if $\mathbf{A'.A} = 1$

With the help of an arbitrarily chosen vector $\mathbf{G}$, we define reciprocals of $\mathbf{A}$ as

$$\mathbf{A'}_{\pm} = \frac{\mathbf{G} \pm \mathbf{iAxG}}{\mathbf{A.G}} \tag{2}$$

We now have the symmetry relation

$$\mathbf{A'}_{+} \hat{+} \mathbf{B'}_{-} = \mathbf{A} \hat{+} \mathbf{B} \tag{3}$$

We intend to study the relation of reflection symmetric sum to Lorentz invariance and also its rotational (spin) property.

## 2. Pauli Quaternion

We construct a 4-dimensional vector [2] and follow the convention adopted by Kyrala [3] to write it as a sum of a scalar and a Cartesian vector

$$\boldsymbol{\mathcal{A}} = A_0 + \mathbf{A}$$

(4)

with the help of basis vectors $\sigma_0$, $\sigma_x$, $\sigma_y$, $\sigma_z$ (they are slightly different from those of Kyrala [3])we write it as

$$\mathbf{A} = \sigma_0 A_0 + \sigma_x A_x + \sigma_y A_y + \sigma_z A_z = \sigma_0 A_0 + \boldsymbol{\sigma}.\mathbf{A} \quad (5)$$

where $A_0$ is a number and $A_x, A_y, A_z$ are the components of the Cartesian vector $\mathbf{A}$

As a consequence of linearity we have

$$k\mathbf{A} = \sigma_0 k A_0 + \boldsymbol{\sigma}.k\mathbf{A} \quad . \quad (6)$$

where $k$ is a constant

From $\mathbf{A} = A_0 + \mathbf{A}$ we construct

$$\mathbf{A}^* = A_0 - \mathbf{A} \quad (7)$$

and define $\mathbf{A}^2 = \mathbf{A}\mathbf{A}^*$ where, so that

$$\mathbf{A}^2 = (\sigma_0 A_0 + \sigma_x A_x + \sigma_y A_y + \sigma_z A_z)\ (\sigma_0 A_0 - \sigma_x A_x - \sigma_y A_y - \sigma_z A_z) \quad (8)$$

or

$$\mathbf{A}^2 = \sigma_0{}^2 A_0{}^2 - \sigma_x^2 A_x{}^2 - \sigma_y^2 A_y{}^2 - \sigma_z^2 A_z{}^2$$
$$(\sigma_x\sigma_0 - \sigma_0\sigma_x)A_0A_x + (\sigma_y\sigma_0 - \sigma_0\sigma_y)A_0A_y + (\sigma_z\sigma_0 - \sigma_0\sigma_z)A_0A_z \quad (9)$$
$$-(\sigma_x\sigma_y + \sigma_y\sigma_x)A_xA_y - (\sigma_x\sigma_z + \sigma_z\sigma_x)A_xA_z - (\sigma_y\sigma_z + \sigma_z\sigma_y)A_yA_z$$

We require

$$\mathbf{A}^2 = A_0{}^2 - \mathbf{A}.\mathbf{A} = A_0{}^2 - \mathbf{A}^2 \quad (10)$$

This will be fulfilled if

$$\sigma_0^2 = \sigma_x^2 = \sigma_y^2 = \sigma_z^2 = 1 \quad (11)$$

$$\sigma_x\sigma_0 - \sigma_0\sigma_x = \sigma_y\sigma_0 - \sigma_0\sigma_y = \sigma_z\sigma_0 - \sigma_0\sigma_z = 0 \quad (12)$$

$$\sigma_x\sigma_y + \sigma_y\sigma_x = \sigma_x\sigma_z + \sigma_z\sigma_x = \sigma_y\sigma_z + \sigma_z\sigma_y = 0 \quad (13)$$

Therefore

$$\sigma_0 = \pm 1 \quad (14)$$

we shall choose

$$\sigma_0 = +1 \quad (15)$$

We require that multiplication of these 4-vectors form a group so that

$$\mathbf{AB} = \mathbf{D} = \sigma_0 D_0 + \boldsymbol{\sigma}.\mathbf{D} = \sigma_0 D_0 + \sigma_x D_x + \sigma_y D_y + \sigma_z D_z \quad (16)$$

where

$$\mathbf{AB} = (\sigma_0 A_0 + \boldsymbol{\sigma}.\mathbf{A})(\sigma_0 B_0 + \boldsymbol{\sigma}.\mathbf{B}) \quad (17)$$

Comparison between (16) and (17) using (11) through (15) shows that we have to have

$$D_0 = A_0 B_0 + \mathbf{A}.\mathbf{B}. \quad (18)$$

$$\boldsymbol{\sigma}.\mathbf{D} = A_0\boldsymbol{\sigma}.\mathbf{B} + B_0\boldsymbol{\sigma}.\mathbf{A}$$
$$+\sigma_x\sigma_y A_x B_y + \sigma_y\sigma_x A_y B_x + \sigma_x\sigma_z A_x B_z \quad (19)$$
$$+\sigma_z\sigma_x A_z B_x + \sigma_y\sigma_z A_y B_z + \sigma_z\sigma_y A_z B_y$$

Group requirement will be fulfilled if we set

$$\sigma_x\sigma_y = -\sigma_y\sigma_x = \beta\sigma_z$$
$$-\sigma_x\sigma_z = \sigma_z\sigma_x = \beta\sigma_y \quad (20)$$
$$\sigma_y\sigma_z = -\sigma_z\sigma_y = \beta\sigma_x$$

where β is a number.

**Associativity**

We also require the operation to be associative. This will be guaranteed if the 'σ' s are associative i.e.

$$(\sigma_i\sigma_j)\sigma_k = \sigma_i(\sigma_j\sigma_k) \qquad for \\ i = 0, x, y, z; \quad j = 0, x, y, z \quad k = 0, x, y, z \tag{21}$$

Then we have

$$\boldsymbol{(AB)C} = \boldsymbol{A(BC)} \tag{22}$$

**To determine β**

Choose relation (20)

$$\sigma_{\mathrm{x}}\sigma_{\mathrm{y}} = \beta\sigma_{\mathrm{z}} \tag{23}$$

Multiplying both sides and using (11)

$$(\sigma_{\mathrm{x}}\sigma_{\mathrm{y}})(\sigma_{\mathrm{x}}\sigma_{\mathrm{y}}) = \beta^2\sigma_{\mathrm{z}}^2 = \beta^2 \tag{24}$$

Using (20) and (21) the above relation becomes

$$-(\sigma_{\mathrm{x}}\sigma_{\mathrm{y}}\sigma_y\sigma_x) = -\sigma_{\mathrm{x}}\sigma_{\mathrm{y}}^2\sigma_{\mathrm{x}} = -\sigma_{\mathrm{x}}^2\sigma_{\mathrm{y}}^2 = -1 = \beta^2 \tag{25}$$

0r

$$\beta^2 = -1 \text{ or } \beta = \pm i \tag{26}$$

We shall choose $\beta = i$

With this (19) becomes

$$\boldsymbol{\sigma}.\mathbf{D} = \boldsymbol{\sigma}.A_0\mathbf{B} + \boldsymbol{\sigma}.B_0\mathbf{A} \\ + i\sigma_z(A_xB_y - A_yB_x) - i\sigma_y(A_xB_z - A_zB_x) + i\sigma_x(A_yB_z - A_zB_y) \tag{27}$$

Or

$$\boldsymbol{\sigma}.\mathbf{D} = \boldsymbol{\sigma}.A_0\mathbf{B} + \boldsymbol{\sigma}.B_0\mathbf{A} + \mathbf{i}\boldsymbol{\sigma}.(\mathbf{AxB}) \tag{28}$$

Or

$$\mathbf{D} = A_0\mathbf{B} + B_0\mathbf{A} + i(\mathbf{AxB}) \tag{29}$$

## 3. Conjugate, Norm and Inverse

$\boldsymbol{A^*}$, the conjugate of $\boldsymbol{A}$, is constructed from $\boldsymbol{A}$ by changing the sign of the Cartesian part of $\boldsymbol{A}$.

$\boldsymbol{A}^2$, the square Norm of $\boldsymbol{A}$, is

$$\boldsymbol{A}^2 = \boldsymbol{AA^*} = \boldsymbol{A^*A} = A_0^2 - \mathbf{A}^2 \tag{30}$$

$\boldsymbol{A}^{-1}$, the inverse of $\boldsymbol{A}$, is

$$\boldsymbol{A}^{-1} = \frac{\boldsymbol{A^*}}{\boldsymbol{A}^2} \tag{31}$$

so that

$$\boldsymbol{A}^{-1}\boldsymbol{A} = \boldsymbol{A}\boldsymbol{A}^{-1} = 1 \tag{32}$$

and the unit vector is 1

Let

$$f(\sigma_x, \sigma_y) = \sigma_x + \sigma_y + \sigma_x\sigma_y \tag{33}$$

Then as a consequence of (16)

$$f(\sigma_x, \sigma_y) = -f(-\sigma_y, -\sigma_x) \tag{34}$$

The product $\mathbf{AB}$ involves quantities of type $f$ above. Therefore,

$$(\mathbf{AB})^* = \mathbf{B}^* \mathbf{A}^* \tag{35}$$

$$(\mathbf{AB})^{-1} = \mathbf{B}^{-1}\mathbf{A}^{-1} \tag{36}$$

$$(\mathbf{AB})^2 = (\mathbf{AB})(\mathbf{AB})^* = \mathbf{ABB}^* \mathbf{A}^* = \mathbf{A}(\mathbf{B}^2)\mathbf{A}^* = \mathbf{A}^2\mathbf{B}^2 \tag{37}$$

## 4. Lorentz Invariant Reflection Symmetric Transformation

Let

$$\mathbf{L} = ct + \mathbf{X} \tag{38}$$

So that, by (30)

$$\mathbf{L}^2 = \mathbf{LL}^* = (ct)^2 - \mathbf{X}^2 \tag{39}$$

and

$$\mathbf{R} = g - g\boldsymbol{\sigma}.\mathbf{V}/c \tag{40}$$

where

$$g = \frac{1}{\sqrt{1-(\mathbf{V}/c)^2}} \tag{41}$$

so that

$$\mathbf{R}^2 = R_0^2 - \mathbf{R}^2 = 1 \tag{42}$$

and

$$\mathbf{RL} = \mathbf{L}' = ct' + \mathbf{X}' \tag{43}$$

With

$$t' = gt - g\mathbf{X}.\mathbf{V}/c^2 \tag{44}$$

and

$$\mathbf{X}' = g(\mathbf{X} - t\mathbf{V}) - ig\frac{\mathbf{VxX}}{c} \tag{45}$$

Using (37) and (42)

$$(\mathbf{L}')^2 = (\mathbf{RL})^2 = \mathbf{L}^2 \tag{46}$$

Therefore, the transformation $\mathbf{L} \to \mathbf{L}'$, (38) through (46), is Lorentz invariant.

## 5. Comparison between the Rotational Properties of Lorentz–Einstein Transformation and Reflection Symmetry Transformation

We define Lorentz-Einstein product of 4-vectors $\mathbf{L}$ and $\mathbf{R}$ defined by (38) and (39) as

$$\mathbf{R}\tilde{\mathbf{x}}\mathbf{L} = cg\{t - \mathbf{X}.\mathbf{V}/c^2\} + \left\{\mathbf{X} + [(g-1)\frac{\mathbf{X}.\mathbf{V}}{\mathbf{V}^2} - gt]\mathbf{V}\right\} \tag{47}$$

Observing that in the limit $c \to 0$

$$1/g = \sqrt{1-(\mathbf{V}/c)^2} \xrightarrow[c \to 0]{} i\frac{|\mathbf{V}|}{c} \tag{48}$$

We find in the limit $c \to 0$

$$\frac{-c\mathbf{R}\tilde{\mathbf{x}}\mathbf{L}}{g|\mathbf{X}||\mathbf{V}|} \xrightarrow[c \to 0]{} \cos\theta + i\frac{1}{|\mathbf{X}|}\left\{\frac{(\mathbf{X}.\mathbf{V})\mathbf{V}}{\mathbf{V}^2} - \mathbf{X}\right\} = \cos\theta + i\mathbf{m}\sin\theta \tag{49}$$

where $\theta$ is the angle between $\mathbf{X}$ and $\mathbf{V}$
Similarly using (43)-(45)

$$\frac{-c\mathsf{RL}}{g\,|\,\mathbf{X}\,||\,\mathbf{V}\,|}\xrightarrow[c\to 0]{}\cos\theta + i\mathbf{n}\sin\theta \tag{50}$$

where

$$\mathbf{m} = \frac{\{(\mathbf{X.V})\mathbf{V}/\mathbf{V}^2\} - \mathbf{X}}{|\,(\mathbf{X.V})\mathbf{V}/\mathbf{V}^2 - \mathbf{X}\,|} \text{ and } \mathbf{n} = \frac{\mathbf{VxX}}{|\,\mathbf{VxX}\,|} \tag{51}$$

An important difference between (49) and (50) is that **m** and **n** are orthogonal so that **m.n** = 0

## 6. Matrix Representation and Spin

For a comparison with the spin of Dirac electron theory, it is worthwhile representing reflection symmetric transformation in matrix formalism.
Pauli matrices [4] have properties (7) and (13). Therefore, it is possible to represented $\sigma_x, \sigma_y$ $\sigma_z$ and $\sigma_0$ by 2x2 Pauli matrices and a unit matrix

$$\sigma_x = \begin{bmatrix} 0 & 1 \\ 1 & 0 \end{bmatrix},\ \sigma_y = \begin{bmatrix} 0 & -i \\ i & 0 \end{bmatrix},\ \sigma_z = \begin{bmatrix} 1 & 0 \\ 0 & -1 \end{bmatrix} \text{ and } \sigma_0 = \begin{bmatrix} 1 & 0 \\ 0 & 1 \end{bmatrix} \tag{52}$$

We now construct the matrix $\mathsf{A}$

$$\mathsf{A} = \sigma_0 A_0 + \sigma_x A_x + \sigma_y A_y + \sigma_z A_z \tag{53}$$

Using

$$\sigma_0 A_0 = \begin{bmatrix} A_0 & 0 \\ 0 & A_0 \end{bmatrix},\ \sigma_x A_x = \begin{bmatrix} 0 & A_x \\ A_x & 0 \end{bmatrix},\ \sigma_y A_y = \begin{bmatrix} 0 & -iA_y \\ iA_y & 0 \end{bmatrix},\ \sigma_z = \begin{bmatrix} A_z & 0 \\ 0 & -A_z \end{bmatrix} \tag{54}$$

we have

$$\mathsf{A} = \sigma_0 A_0 + \boldsymbol{\sigma}.\mathbf{A} = \begin{bmatrix} A_0 + A_z & A_x - iA_y \\ A_x + iA_y & A_0 - A_z \end{bmatrix} \tag{55}$$

Constructing matrices of type (55) using $\mathsf{L}$ and $\mathsf{R}$ of (38) and (40) and multiplying we get
$\mathsf{L}' = \mathsf{RL}$
The product, $\mathsf{L}'$, is a 2x2 matrix. The cross product term of $\mathsf{L}'$, corresponding to cross product term of (45), corresponds to the spin term of Dirac electron. Spin term of Dirac electron has its origin in the product of Pauli matrices [5].

## 7. Normalized Vector and Addition of Velocities

The vector $\mathsf{W}$ will be called normalized for velocity when its scalar part is *c;* i.e. $W_0 = c$ In this case its Cartesian part will be called a velocity.
Let

$$\mathsf{U} = U_0 + \mathbf{U} \tag{56}$$

Consider the product

$$\mathsf{RU} = \mathsf{W}' = \frac{R_0 U_0 + \mathbf{R.U}}{c}\mathsf{W} \tag{57}$$

$$\mathsf{W} = c + \boldsymbol{\sigma}.\mathbf{W} \tag{58}$$

We set $U_0 = c$. $\mathsf{R}$ is as defined by (38). **U** and **V** are velocities and

$$\mathsf{RU} = \mathsf{W'} = \frac{1-\mathbf{V}.\mathbf{U}/c^2}{\sqrt{1-(\mathbf{V}/c)^2}}\mathsf{W} \tag{59}$$

$$\mathsf{W} = W_0 + \boldsymbol{\sigma}.\mathbf{W} = c + \boldsymbol{\sigma}.\frac{-\mathbf{V} + \mathbf{U} - \mathbf{i}\frac{\mathbf{1}}{c}\mathbf{VxU}}{\mathbf{1} - \frac{\mathbf{V.U}}{c^2}} \tag{60}$$

We introduce the symbol $\hat{+}$ and write

$$\mathbf{W} = (\mathbf{-V})\,\hat{+}\,\mathbf{U} = \frac{-\mathbf{V} + \mathbf{U} - \mathbf{i}\frac{\mathbf{1}}{c}\mathbf{VxU}}{\mathbf{1} - \frac{\mathbf{V.U}}{c^2}} \tag{61}$$

As a consequence of (29), (48) is associative

$$\{(-\mathbf{V})\,\hat{+}\,\mathbf{U}\}\,\hat{+}\,\mathbf{W} = (-\mathbf{V})\,\hat{+}\,\{\mathbf{U}\,\hat{+}\,\mathbf{W}\} \tag{62}$$

## 8. Comparison with Lorentz-Einstein Transformation

### The magnitude

The magnitude is the same in both the cases, reflection symmetric, ($\hat{+}$) and L-E, ($\tilde{+}$).

$$|\,\mathbf{V}\,\hat{+}\,\mathbf{U}\,| = |\,\mathbf{V}\,\tilde{+}\,\mathbf{U}\,| \tag{63}$$

### Invariance of Limiting Velocity under Reflection Symmetric Addition

We observe that

$$\mathbf{A}\,\hat{+}\,\mathbf{G} \xrightarrow[\mathbf{G}\to\infty]{} \mathbf{A'}_{+} = \frac{\mathbf{G} + \mathbf{iAxG}}{\mathbf{A.G}} \tag{64}$$

Let $\mathbf{W}$ be

$$\mathbf{W} = \mathbf{A}\,\hat{+}\,\mathbf{B} \tag{65}$$

Because of associativity

$$\mathbf{W}\,\hat{+}\,\mathbf{G} = \mathbf{A}\,\hat{+}\,(\mathbf{B}\,\hat{+}\,\mathbf{G}) \tag{66}$$

Going to the limit $\mathbf{G} \to \infty$

$$\mathbf{W'} = \mathbf{A}\,\hat{+}\,\mathbf{B'} \tag{67}$$

By (2)

$$\mathbf{W.W'} = 1 \tag{68}$$

Therefore

$$(\mathbf{A}\,\hat{+}\,\mathbf{B}).(\mathbf{A}\,\hat{+}\,\mathbf{B'}) = \mathbf{1} \text{ if } \mathbf{B.B'} = \mathbf{1} \tag{69}$$

If

$$\mathbf{B} = \mathbf{B'} \tag{70}$$

corresponding to the choice

$$\mathbf{G} = \mathbf{V} \text{ and } |\,\mathbf{V}\,| = 1 \tag{71}$$

Then (72) becomes

$$\mathbf{W.W} = (\mathbf{A}\,\hat{+}\,\mathbf{B}).(\mathbf{A}\,\hat{+}\,\mathbf{B}) = \mathbf{1} \text{ if } \mathbf{B.B} = \mathbf{1} \tag{72}$$

We may replace all the quantities **A, B** and **W** by

$$\mathbf{A} \to \mathbf{a}/c, \mathbf{B} \to \mathbf{b}/c \;\; and \;\; \mathbf{W} \to \mathbf{w}/c \tag{73}$$

Then (72) gives

$$\mathbf{w.w} = c^2 \text{ if } \mathbf{b.b} = c^2 \tag{74}$$

**a, b** and **w** have the dimension of $c$ and **A, B** and **W** are dimensionless. If $c$ is a velocity, the magnitude of the sum of velocities **a** and **b** is also c. This is the reflection symmetric analogue of Einstein's postulate.

## 9. Conclusion

A non-commutative sum with reflection property (3) has been written. It is Lorentz invariant (46) and associative (22, 29). It has a spin-like property comparable to that of Dirac electron (sections 6). Lorentz transformation also has this spin-like property (49) but is hidden and becomes manifest in the limit $c \to 0$. Relation (74) corresponds to Einstein's postulate. We have found it as a consequence of reflection symmetry.

## References

[1]. Mushfiq Ahmad, M. Shah Alam, O. G. Talukder. Reflection Symmetric Number System And Correspondence Between Quantum Mechanical And Classical Quantities. Submitted to IJTP. IJTP-468.

[2] George Raetz. http://home.pcisys.net/~bestwork.1/quaterni2.html.

[3]. A. Kyrala. Theoretical Physics: Applications of Vectors, Matrices, Tensors and Quaternions. W. B. Saunders Company, Philadelphia & London. 1967

[4] Schiff, L.I (1970). Quantum Mechanics. Mc Graw-Hill Company. P.84.

[5] Schiff, L.I (1970). Quantum Mechanics. Mc Graw-Hill Company. P.84.